\documentclass[a4paper]{jpconf}
\usepackage{graphicx}
\begin{document}
\title{Two Approaches to Testing General Relativity in the 
Strong-Field Regime}

\author{Dimitrios Psaltis}

\address{Departments of Astronomy and Physics, University of
Arizona, 933 N.\ Cherry Ave., Tucson, AZ 85721}

\ead{dpsaltis@email.arizona.edu}

\begin{abstract}
Observations of compact objects in the electromagnetic spectrum and the
detection of gravitational waves from them can lead to quantitative
tests of the theory of general relativity in the strong-field regime
following two very different approaches. In the first approach, the
general relativistic field equations are modified at a fundamental
level and the magnitudes of the potential deviations are constrained
by comparison with observations.  In the second approach, the exterior
spacetimes of compact objects are parametrized in a phenomenological
way, the various parameters are measured observationally, and the
results are finally compared against the general relativistic
predictions. In this article, I discuss the current status of both
approaches, focusing on the lessons learned from a large number of
recent investigations.
\end{abstract}

\section{Introduction}

Testing the theory of general relativity is the key science objective
of many future astrophysical missions as well as of NASA's Beyond
Einstein program, as a whole. However, there has been very little
consensus, as of now, on the particular direction along which
quantitative tests of the theory will be performed.

The parametric post-Newtonian framework, which has been crucial in
testing general relativity at solar-system scales, is no longer valid
in the strong-field regime. In fact, in the absence of an accepted
alternative theory of gravity, there are very few a priori constrains
on the potential deviations from general relativistic
predictions. However, there are a number of arguments that may guide
us in the effort to reach this goal.

As in most other branches of physics, there are two distinct avenues
we may follow in testing general relativity in the strong-field
regime. In a top-down approach, we may modify the theory at a
fundamental level and calculate the consequences of the modifications
that can be tested observationally. In a bottom-up approach, we will
use a phenomenological description of the observations in order to
obtain clues for how to modify the theory at the fundamental level. In
this article, I draw a roadmap for testing the theory of general
relativity in the strong-field regime along these two avenues, based
on general theoretical expectation as well as on a large number of
recent investigations.

\section{The top-down approach: From Theory to Observations}

\bigskip

\subsection{The Equivalence Principle and the Einstein Field Equations}

The theory of general relativity has two distinct ingredients. The
equivalence principle that describes how matter moves in the presence of a
gravitational field and the Einstein field equations that describes how the
gravitational field is generated in the presence of matter. 

The validity of the equivalence principle gives rise to the geometric
aspect of the theory (see Will 2006). According to this principle, it
is impossible to tell the difference between a reference frame at rest
and one free-falling in a gravitational field, by performing local
experiments.  Moreover, the equivalence principle encompasses the
Lorentz symmetry, as well as our beliefs that there is no preferred
frame and position anywhere in the universe.

The Einstein equivalence principle has been tested to a very high
degree during the last century, mostly in the weak-field regime. The
upper bound on possible violations are as low as one part in $10^{12}$
(Will 2006). It is very unlikely that astrophysical observations of
black holes and neutron stars can lead to tests of comparable
accuracy. For this reason, most approaches to date have concentrated
only on plausible modifications of the field equations although
frameworks (such as the Standard Model Extension; see Kostelecky
2003 and references therein) exist that can lead, in principle,
to strong-field tests of this principle (see Stairs 2003 for a
discussion of testing the strong equivalence principle with double
neutron stars; see also Eling \& Jacobson 2006 for black-hole solutions
in an Einstein-aether theory).

\subsection{Modifying the Einstein Field Equations}

\begin{figure}[t]
\begin{center}
\includegraphics[height=15cm]{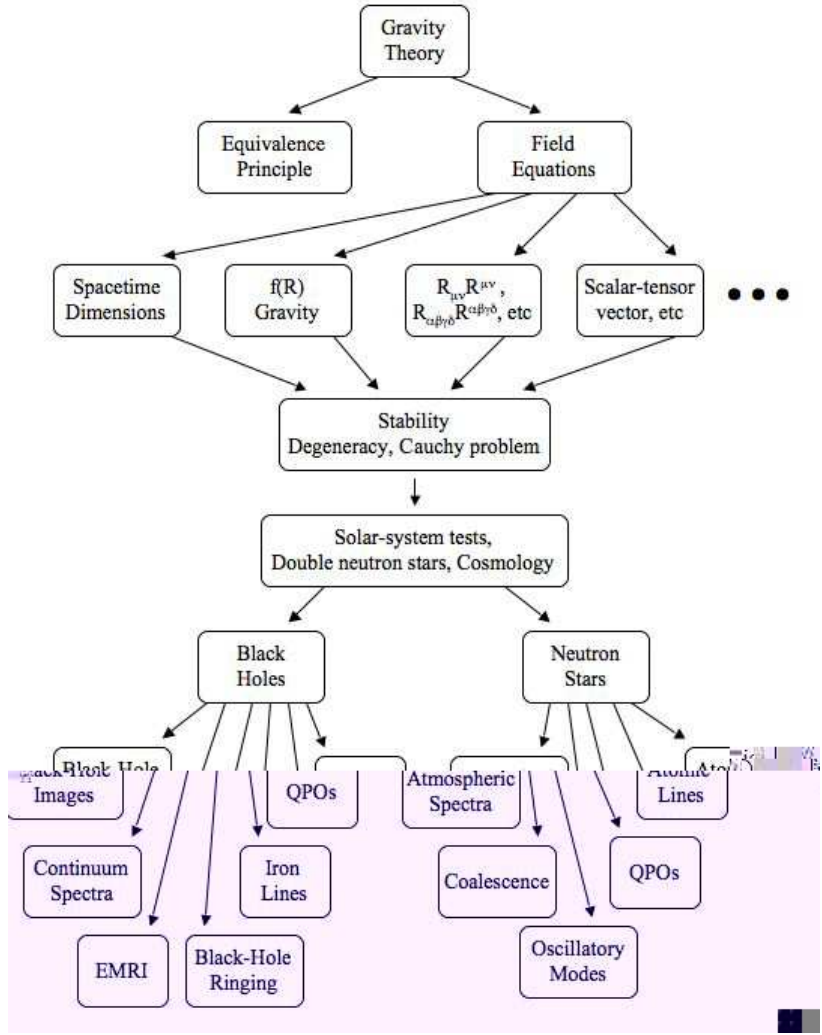}
\end{center}
\caption{\label{figure1}A top-down approach to performing strong-field
tests of gravity.}
\end{figure}

Contrary to the case of the equivalence principle, there are no
arguments one can make that lead uniquely to the Einstein field
equations. Indeed, Einstein reached the field equation, more or less,
by reverse engineering. For this reason, his approach cannot be easily
modified to include additional terms. On the other hand, the
``derivation'' of the field equations from a Lagrangian action
provided by Hilbert allow us to modify the theory, while preserving
many of its symmetries and conservation laws.

The Einstein-Hilbert action that leads to the field equations of
general relativity is directly proportional to the Ricci scalar, $R$,
\begin{equation}
           S= \frac{c^4}{16\pi G}\int d^4 x  \, \sqrt{-g} \, (R-2\Lambda)\;,
\label{EHaction}
\end{equation}
where $g$ is the spacetime metric, $c$ is the speed of light, $G$ is
the gravitational constant, and $\Lambda$ is the cosmological
constant.  

We can construct a self-consistent theory of gravity from this
starting point using any other action that obeys the following four
simple requirements (see Misner et al.\ 1973; p.~410). It has to: {\em
(i)\/} reproduce the Minkowski spacetime in the absence of matter and
cosmological constants, {\em (ii)\/} be constructed from only the
Riemann curvature tensor and the metric, {\em (iii)\/} follow the
symmetries and conservation laws of the stress-energy tensor of
matter, and {\em (iv)\/} reproduce Poisson's equation in the Newtonian
limit. Of all the possibilities that meet these requirements, the
field equations that are derived from the Einstein-Hilbert action are
the only ones that are also linear in the Riemann tensor. 

The function form of the Einstein-Hilbert action suggests that we can
modify it in one of four ways: In principle, we can {\em (i)\/} change
the number of spacetime dimensions, {\em (ii)} change the functional
dependence of the Lagrangian density on the Ricci scalar $R$, {\em
(iii)} allow for the Lagrangian density to depend on other scalars
generated from the Riemann curvature tensor, or {\em (iv)\/} introduce
additional scalar, vector, or tensor fields.

\bigskip

\noindent {\em (i) The number of spacetime dimensions.---\/}The
possibility that we live in a Universe with more than four spacetime
dimensions has been the subject of intense research during the last
decade, primarily as a solution to the hierarchy problem in physics
(Arkani-Hamed et al.\ 1998a; Randall \& Sundrum 1999; Maartens
2005). In these braneworld gravity models, all non-gravitational
fields and particles live on a four dimensional spacetime, whereas the
gravitational field is allowed to propagate in the extra
dimensions. In order for the theory to be consistent with Newton's law
of gravity down to the sub-mm level, the effective size of the extra
dimensions has to be smaller than $\sim 50~\mu$m (Kapner et al.\
2007). As a result, such a modification to gravity is not expected to
have any observable classical consequences even for neutron stars,
which are the smallest known astrophysical objects with strong
gravitational fields (Arkani-Hamed et al.\ 1998b).

A plausible consequence of one variant of braneworld gravity (the RS2
scenario of Randall \& Sundrum) that can be constrained with
astrophysical observations is the rapid evaporation of black holes
caused by the emission of gravitons in the extra dimensions (Emparan
et al.\ 2003 and references therein). The rate of rapid evaporation
was inferred using the AdS/CFT correspondence and is a matter of debate
(Fitzpatrick et al.\ 2006). If proven accurate, however, it can lead
to astrophysical bounds on the asymptotic scale of the large extra
dimensions that are comparable to those obtained in table-top
experiments (Johannsen et al.\ 2009 and references therein).

\bigskip

\noindent {\em (ii) The functional dependence of the Langrangian 
density on the Ricci scalar.---\/}
A Langrangian action that depends on a non-linear function of the 
Ricci scalar $R$, i.e.,
\begin{equation}
           S =\frac{c^4}{16\pi G}\
                 \int d^4 x  \, \sqrt{-g} \,
            f(R)\;,
\label{EHaction2}
\end{equation}
leads to field equations that obey all the requirements outlined
above.  Indeed, the action~(\ref{EHaction2}) results in a field
equation that allows for the Minkowski solution in the absence of
matter, is constructed only from the Riemann tensor, obeys the usual
symmetries and conservation laws, and can be made to produce
negligible corrections at the small curvatures probed by weak-field
gravitational experiments.  Such non-linear actions for the
gravitational field, albeit with negligibly small coefficients, occur
naturally in quantum gravity theories and in string theory. This is
true because of radiative corrections to the usual Einstein-Hilbert
action that induce an infinite series of counterterms that cannot be
reabsorbed into the original Lagrangian by adjusting its bare
parameters (see discussion in Burgess 2004).

Despite their simplicity and appeal, a number of important limitations
to $f(R)$ gravity theories have been recently explored in the
cosmological context (see Sotiriou \& Faraoni 2009 for an extensive
review). Modifications to the Lagrangian action that appear negligible
at solar-system scales may lead to order unity deviations of the PPN
parameters that are inconsistent with current experiments (Chiba
2003). Moreover, astrophysical objects in non-linear $f(R)$ gravity,
including the Universe as a whole, can be violently unstable,
depending on the sign of $d^2f/dR^2$ (Dolgov \& Kawasaki 2003).
Finally, special attention has to be paid to ensure that a well posed
initial value problem can be defined under all physical conditions
within the non-linear $f(R)$ theory (Lanahan-Tremblay \& Faraoni
2007).

Resolving the above problems requires a carefully constructed $f(R)$
theory with fine-tuned coefficients of the non-linear terms (e.g.,
Nojiri \& Odintsov 2003) or one that leads to the so-called chameleon
mechanism (see Hu \& Sawicki 2007 and references
therein). Alternatively, a simple $f(R)$ function may be considered as
an expansion of an unknown, more general theory and its consequences
for astrophysical phenomena may be calculated within the context of
perturbative constraints (Simon 1990; see also Cooney et al.\ 2009 and
references therein).

In adding terms to the Lagrangian action that are of non-linear order
in the curvature, we have the choice of performing the derivation of
the parametric Einstein equation either in the metric formalism or in
the Palatini formalism. The first corresponds to extremizing the
action under variations in the metric only, whereas the second
corresponds to extremizing the action under variations in the metric
and the connection, as well. For the simple Einstein-Hilbert action,
both approaches are equivalent and give rise to the same field
equations. However, when the action has non-linear terms in the Ricci
scalar, the two approaches diverge. Recent studies of $f(R)$ gravity
in the Palatini formalism, however, have revealed a number of serious
problems with this approach that are related to the suppression of the
new dynamical degrees of freedom and render it a rather unlikely
alternative to the usual derivation of the field equations from a
Lagrangian action (see Sotiriou \& Faraoni 2009 and references
therein for a detailed discussion).

\bigskip

\noindent  {\em (iii) Additional scalar terms 
constructed from the Riemann curvature tensor.---\/} It is remarkable
that the Einstein field equations can be derived from a first-order
Lagrangian action that does not involve the Riemann tensor (see Landau
\& Lifshitz 1992). However, higher-order field equations that obey the Lorenz
symmetry have been derived only from Lagrangian actions that involve
different scalar quantities constructed from the Riemann tensor
$R_{\alpha\beta\gamma\delta}$, e.g.,
\begin{equation}
{\cal S}= \frac{c^4}{16\pi G}\int d^4x
   \sqrt{-g}\left(R+\alpha R^2 + \beta R_{\sigma \tau}R^{\sigma \tau}
   + \gamma R^{\alpha\beta\gamma\delta} R_{\alpha\beta\gamma\delta}\right)\;,
   \label{eq:PL}
\end{equation}
where $R_{\sigma\tau}$ is the Ricci tensor and $\alpha$, $\beta$, and
$\gamma$ the three parameters of the theory

Similar to the case of $f(R)$ gravity, such terms arise naturally as
high-order corrections in quantum gravity and string theory and their
relative importance increases with the curvature of the metric.  The
resulting field equations suffer, however, from the Ostragardski
instability, which can be avoided only by fine tuning of the free
parameters of the theory (see Woodard 2006 and references
therein). 

\bigskip

\noindent {\em (iv) The presence of additional gravitational
fields.---\/}The single, rank-2 tensor field $g_{\mu\nu}$ (i.e., the
metric) of the Einstein-Hilbert action may also not be adequate to
describe completely the gravitational force (although, if additional
fields are introduced, then the strong equivalence principle is
violated, with important implications for the frame- and time-
dependence of gravitational experiments). 

The simplest case of a theory with additional gravitational fields
involves a single additional scalar and has been studied for more than
40 years in the form of Brans-Dicke gravity (e.g., Will 2006). Because
of its particular properties, the fractional deviation of the
Brans-Dicke predictions from general relativity are comparable in both
the weak and strong-field regimes. In a more general case, however,
depending on the coupling between the metric, the scalar field, and
matter, the relative contribution of such additional fields may become
significant only at the high curvatures found in the vicinity of
compact objects.

The general form of the Lagrangian of a scalar-tensor theory is given, 
in the Einstein frame, by the Bregmann-Wagoner action (see Will 2006 for
details)
\begin{equation}
S=\frac{c^4}{16\pi G}\int d^4x \sqrt{-g_*}\left[R_* \pm 
g_*^{\mu\nu}\partial_\mu \phi \partial_\nu\phi +
2\lambda(\phi)\right]+S_{\rm m}[\phi_m,A^2(\phi)g_{\mu\nu}]\;,
\label{eq:st}
\end{equation}
where $A(\phi)$ and $\lambda(\phi)$ are two arbitrary functions, and
$S_m$ is the action for the matter field $\phi_m$.  In the
strong-field regime, the potential term $\lambda(\phi)$ in the
action~(\ref{eq:st}) is typically negligible. The single coupling
function $A(\phi)$ encodes the properties of the theory and may be
parametrized as $A(\phi)= exp[\phi_0+\alpha_\phi
\phi+(1/2)\beta_\phi \phi^2]$, with $\phi_0\rightarrow 0$ 
being the cosmological value of the field (Damour \& Esposito-Farese
1993).

\subsection{Studying the Properties of a Parametric Lagrangian Theory}

The parametric addition of terms in the Einstein-Hilbert action offers
a systematic and self-consistent way of characterizing potential
deviations from general relativistic predictions that can be
constructed to become arbitrarily large in the strong-field regime. It
may also lead, however, to a number of pitfalls that need to be
studied and understood before embarking onto a detailed comparison to
observational data.

First, a parametric extension to the Einstein-Hilbert action is guided
only by the mathematical properties of the form that gave rise to the
field equations of general relativity and not by a new requirement of
the physics.  As such, the extensions discussed above will not cover
the entire range of possibilities, let alone a theory that is
fundamentally different from general relativity in the strong-field
regime. In some sense, using the parametric extension of the
Einstein-Hilbert action will allow us to look for deviations from
general relativity albeit not necessarily the ones that are described
by the particular parameters measured to differ from their general
relativistic values.  If future experiments do measure such a
deviation, then a more fundamental study will be required to account
for it.  There is a close similarity between this situation and the
current measurements of the cosmological constant in the very weak
field regime. Indeed, the non-zero measured value of the cosmological
constant may just be an indication for the presence of a dynamical
scalar field (such as quintessence) as opposed to a constant in the
Einstein field equations. However, until a non-zero cosmological constant was
measured, it made little sense to introduce the additional
complication of a dynamical equation for the scalar field in the
Friedman equation. 

Second, taken at face value, the predictions of a parametric
Lagrangian theory may describe unstable physical systems, as discussed
above. Albeit challenging to deal with, such theories need not be {\em
a priori\/} excluded from quantitative comparison with astrophysical
data. There have been, indeed, many cases in physics where unstable
solutions have been tested against data (e.g., the cosmological
solution with $\Omega\ne 1$ is highly unstable but is still being
tested against WMAP data) and even some situations in which the
unstable solution was preferred over the stable one (e.g., the
Rutherford model of the atom). When the experimental data prefer a
steady-state solution that is unstable, it is almost always an
indication of new physics, which would have been missed had such a
solution been excluded.

The same argument can also be made for predictions that appear
to be unphysical in the context of the parametric theory but are not
at a more fundamental level. This is similar to some predictions of
the PPN formalism related, e.g., to non-zero values of the parameters
$\alpha_3$, and $\zeta_1$ through $\zeta_4$ that describe violations
of conservation of total momentum.  Measuring any of these
``unphysical'' parameters to be anything other than zero would be
interpreted simply as a smoking gun for the presence of an additional
field that has not been incorporated in the PPN formalism but carries
some momentum away from the system under study.

Finally, it is also worth emphasizing that there is significant
degeneracy between theories that arise from different Lagrangian
actions and, therefore, not every astrophysical observation can
distinguish between them. For example, it is well known that the field
equations of gravity theories with higher-order terms in the Ricci
scalar are completely equivalent to those of a scalar-tensor theory
(see, e.g., Magnano \& Sokolowski 1994) although this is no longer
true when additional terms are introduced to the action that involve
the Riemann and Ricci tensors, i.e., $R^{\mu\nu}R_{\mu\nu}$ and
$R^{\alpha\beta\mu\nu}R_{\alpha\beta\mu\nu}$. Even in the latter case,
however, the particular predictions of the theory for specific
astrophysical systems is not always different from their general relativistic
counterparts.

Consider, for example, the theory given by the Lagrangian
action~(\ref{eq:PL}).  Because of the Gauss-Bonnet identity,
\begin{equation}
\frac{\delta}{\delta g_{\mu\nu}}\int d^4x \left(
   R^2 -4 R_{\sigma \tau}R^{\sigma \tau}
   +  R^{\alpha\beta\gamma\delta} R_{\alpha\beta\gamma\delta}\right)
   =0\;,
\end{equation}
variations of the term proportional to $\gamma$ in equation~(\ref{eq:PL}),
with respect to the metric, can be expressed as variations of the
terms proportional to $\alpha$ and $\beta$. Therefore, for all classical
tests, the predictions of the theory described by the Lagrangian
action~(\ref{eq:PL}) are identical to those of the Lagrangian
\begin{equation}
{\cal S}= \frac{c^4}{16\pi G}\int d^4x
   \sqrt{-g}\left[R+(\alpha-\gamma) R^2 + (\beta+4\gamma) 
    R_{\sigma \tau}R^{\sigma \tau}\right]\;.
   \label{eq:r2_clas}
\end{equation}
As a result, classical astrophysical tests can only constrain a 
particular combination of the parameters, i.e., $\alpha-\gamma$ and 
$\beta+4\gamma$. It is only through phenomena related to quantum gravity, 
such as the evaporation timescale of black holes, that the parameter 
$\gamma$ may be constrained.

When the spacetime is isotropic and homogeneous, as in the case of 
cosmological tests, an additional identity is satisfied, i.e.,
\begin{equation}
\frac{\delta}{\delta g_{\mu\nu}}\int d^4x \left(
   R^2 -3 R_{\sigma \tau}R^{\sigma \tau}\right)=0\;.
\end{equation}
This implies that, for cosmological tests, the predictions of the theory
described by the Lagrangian action~(\ref{eq:PL}) are identical to those 
of the Lagrangian
\begin{equation}
{\cal S}= \frac{c^4}{16\pi G}\int d^4x
   \sqrt{-g}\left[R+(\alpha+\frac{1}{3}\beta+\frac{1}{3}\gamma) R^2\right]\;.
   \label{eq:r2_cosm}
\end{equation}
As a result, cosmological tests of gravity can only constrain a
particular combination of the parameters, i.e.,
$\alpha+1/3\beta+1/3\gamma$.

\subsection{Reducing the Size of the Parameter Space of Alternative Theories}

The general relativistic field equations in the presence of matter are
highly non-linear and the same is true for all alternatives explored
so far. Because of this non-linearity, modifications to the theory
that are apparently significant, e.g., only in the strong-field regime
may have observable consequences at very different astrophysical or
cosmological scales and vice versa. Recent studies of $f(R)=R\pm\mu^4/R$
gravity theories have actually provided one of the most
straightforward demonstrations of this effect. Although the term
$\mu^4/R$ in the action appears to become relevant only when the
curvature of the field is $R\le \mu^2$, the PPN parameter $\gamma$ for
such a theory is equal to $1/2$, is independent of the value of the small
parameter $\mu^2$, and is demonstratively different that the general
relativistic prediction of $\gamma=1$ (Chiba 2003). 

Because of this non-linearity in the field equations, the vast parameter
space of alternatives to general relativity can be reduced by
examining their behavior in the weak-field regime or even at cosmological
scales. In particular, every modification to general relativity 
has to account for

\bigskip

\noindent {\em (i) the observed constraints on the PPN 
parameters.---\/} In particular, the PPN parameters $\beta$
and $\gamma$ have been constrained to lie within $\simeq
10^{-5}$ of their general relativistic values using tests involving
bodies in the solar-system as well as man-made spacecraft (Will 2006
and references therein).  Calculations of the PPN parameters in
scalar-tensor theories specifically designed to perform strong-field
tests have been performed by Damour \& Esposito-Farese (1993), whereas
a large number of similar studies have been carried out for $f(R)$
gravity theories (Barrause \& Sotiriou 2009).

\bigskip

\noindent {\em (ii) the orbital period decay of double neutron 
stars.---\/} This indirect evidence for the emission of gravitational
waves offers a unique setting in which the predictions of a gravity
theory for the time dependence of a dynamical spacetime can be
compared against highly accurate observations (Stairs 2003). Damour \&
Esposito-Farese (1993) have exploited these observations to place
strong constraints on strong-field modifications to general relativity
that involve dynamical scalar fields.

\bigskip

\noindent  {\em (iii) the evolution of the Universe predicted 
by the modified Friedman equation}. Reversing the argument given
earlier for cosmologically motivated theories, it is important that a
modification to gravity constructed for use in strong-field tests does
not contradict the constraints imposed by the observations of the
high-z supernovae (which probe the nearby universe, up to a redshift
of $\sim 1$), of the cosmic microwave background (which is sensitive
to the expansion of the universe up to a redshift of $\sim 1000$), and of
big-bang nucleosynthesis (which is highly sensitive to the rate of the
expansion of the universe at a much earlier time). 

\subsection{Black Holes and Neutron Stars in Alternative Gravity Theories}

The final step in setting up tests of strong-field gravity in the
top-bottom approach discussed here involves calculating the
properties of black holes and neutron stars in the parametric Lagrangian
theories.

In the case of neutron stars, the presence of matter in the domain of
solution guarantees that the addition of terms to the Einstein field
equations will lead to modifications in the structure of the stars and
their exterior spacetimes. Indeed, studies of neutron stars in
alternative gravity theories have invariably resulted in larger and
more massive neutron stars compared to their general relativistic
counterparts. Zaglauer (1992) and Damour \& Esposito-Farese (1993)
studied the structure of neutron stars in gravity theories that
involve a parametric addition of a scalar field to the
Einstein-Hilbert action. More recently, a number of studies
investigated compact objects in $f(R)$ gravity and in particular the
role of the chameleon mechanism in allowing neutron stars to exist in
a large class of theories (see Upadhye \& Hu 2009 and references
therein).

Contrary to the case of neutron stars, most parametric extensions to
the Einstein-Hilbert action lead to field equations that admit as a
solution the Kerr spacetime of a black hole (Psaltis et al.\
2008). This is formally true only for steady-state black holes, as the
time-dependent spacetime of a perturbed black hole that can be probed
with observations of gravitational waves is different in different
theories (Barausse \& Sotiriou 2008). The reason behind the similarity
of black hole solutions among widely different theories is the absence
of matter everywhere in the domain of solution. Indeed, the entire
spacetime of a black hole as well as that of vacuum are characterized
by a constant Ricci scalar curvature $R$. As a result, calculating the
spacetime of a black hole requires simply solving the equation
$R=$constant, independent of the underlying field equations of the
theory.

Recently, Yunes \& Pretorius (2009) as well as Konno et al.\ (2009)
showed that, in Chern-Simons gravity, rotating black-hole solutions
are different from the Kerr spacetimes even though the theory admits
the normal vacuum solution. They key difference between Chern-Simons
gravity and general relativity is the addition of a term in the
Einstein-Hilbert action that allows for solutions that violate parity.
The spacetime for the vacuum and for a Schwarzschild black hole have
zero parity and are, therefore, the same in both theories. On the
other hand, the spacetime of a rotating black hole has non-zero parity
and, in Chern-Simons gravity, is parametrically different from the
Kerr solution.

Having obtained the structure of neutron stars and exterior spacetimes
of black holes, the theory can now be put to test against
observations that probe strong gravitational fields. In the
electromagnetic spectrum, such tests can be provided by observations
of the images of black-hole accretion flows, of the continuum and line
spectra of neutron stars and black holes, as well as of their rapid
variability properties (see Psaltis 2008 for a review). In the near
future, gravitational wave observations of coalescing compact objects,
of extreme mass-ratio inspirals, as well as of the ringing of
perturbed black holes and of the global oscillatory modes of neutron
stars will open a new window to the study of the strong-field
properties of gravity (see Flanagan \& Hughes 2005 for a review).

\section{The bottom-up approach: From Phenomenology to Observations}

\begin{figure}[t]
\begin{center}
\includegraphics[height=13cm]{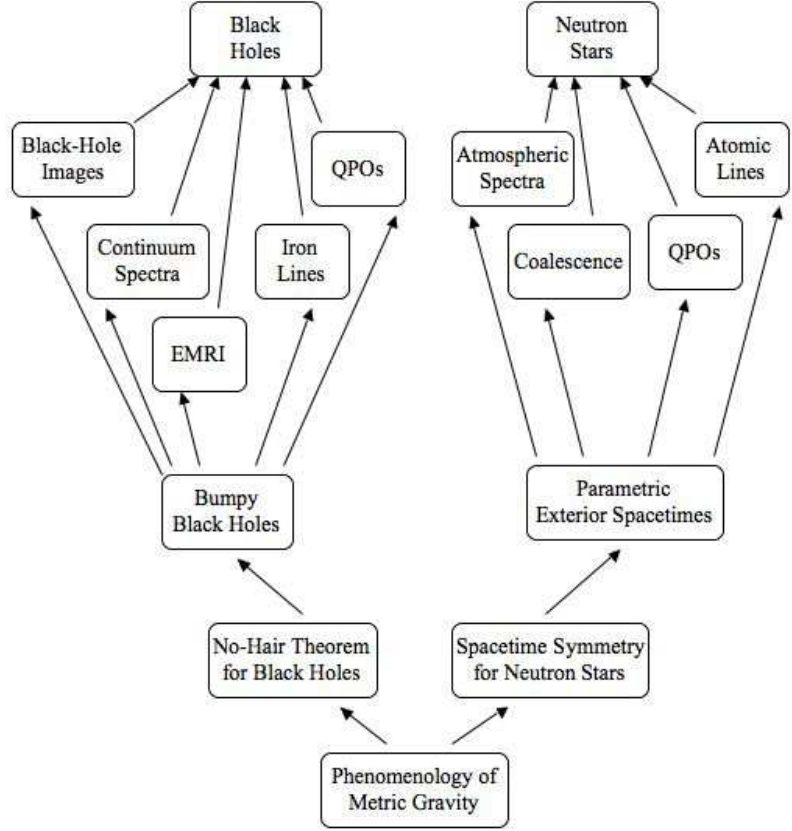}
\end{center}
\caption{\label{figure2}A bottom-up approach to performing strong-field
tests of gravity.}
\end{figure}

The top-down approach to testing the strong-field regime of gravity
discussed above is appealing for a number of reasons. First, the terms
added to the Einstein-Hilbert action are often motivated by quantum
gravity and string theory, offering therefore the possibility of
testing predictions of a fundamental modification to gravity. Second,
starting from a Lagrangian action guarantees that the theory will
encompass important symmetries and conservation laws. Finally, the
parameters of the same theory can be constrained by a wide range of
observations of neutron stars and black holes, in the electromagnetic
spectrum and with gravitational waves, or even with solar-system
experiments and in cosmological settings. As a result, very tight
constraints can be obtained and degeneracies between parameters can be
broken.

The key drawback of the top-down approach, however, is the fact that
observational data are interpreted within the narrow confines of a
particular theoretical framework. The case of black-hole spacetimes
discussed earlier is the most striking example of the limitations of
this approach: a very wide range of gravity theories lead to only one
black-hole solution potentially modified by a parity violating term,
thereby allowing only for a very limited range of possibilities to be
tested against observations.

As an alternative to this approach, we might follow a different,
phenomenologically driven path. We will start again by assuming the
validity of the equivalence principle and hence of the fact that test
particles and photons follow geodesics of the spacetime. In this
approach, however, we will not aim to obtain the spacetime as a
solution to a field equation. Instead, we will prescribe the external
spacetime of a black hole or a neutron star in a way that is
parametrically different from their general relativistic
counterparts. Using these spacetimes we can then predict observational
signatures of deviations from general relativistic predictions to be
tested observationally.  This approach has been followed only
recently and for a limited number of cases, which I will discuss
below.

\subsection{Parametric Spacetimes of Black Holes and Neutron Stars}

The exterior spacetime of an astrophysical black hole in general
relativity is completely described by only two parameters: its mass
and its spin (any charge on the black hole will be very quickly
neutralized by attracting opposite charges from the surrounding
medium). This is the so-called no-hair theorem that provides the
framework on which to build a phenomenological model of black-hole
spacetimes (this approach was pioneered by Ryan 1995 and was further
developed more recently by Collins \& Hughes 2004 and Glampedakis \&
Babak 2006).

Because of the no-hair theorem, if we expand the exterior spacetime of
a black hole in multipole moments, the coefficients of only two of
them are independent quantities: the monopole is related to the
black-hole mass $M$ and the dipole to its spin $a$. All higher-order
moments depend in a particular way on these two. For example, the
dimensionless quadrupole moment of a black hole in general relativity
is equal to $q=-a^2$. We can, therefore, write the most general
spacetime with arbitrary coefficients of its multipole moments and use
it to predict observables such as the spectra of accretion flows
around black holes or the waveforms of gravitational waves during
extreme mass-ratio inspirals. Measuring the coefficients of at least
three multipole moments of a black hole spacetime by comparing these
predictions to observations we can, therefore, test this fundamental
property of general relativistic black-hole spacetimes and search for
potential deviations.

This approach is very powerful for performing tests of strong-field
gravity with black holes and perhaps the one that is the least model
dependent. This is especially true for tests that will use
observations in the electromagnetic spectrum, because the calculation
of the propagation of photons from the strong gravitational fields to
the distant observers does not require the knowledge of the underlying
field equations that gave rise to the parametric spacetimes. In order,
however, to predict the waveforms of gravitational waves emitted, for
example, from an extreme mass-ratio inspiral into such a spacetime one
needs to use a field equation to describe the time-dependence of the
parametric spacetime.  As a result, gravitational wave observations
can be used within this parametric framework only as null hypothesis
tests of general relativistic predictions (see discussion in Hughes
2006).

Contrary to the case of black holes, the exterior spacetimes of
neutron stars depend, in general, on the mass distribution in their
interiors and are not uniquely described by only two coefficients of
their multiple moments. On the other hand, a number of phenomena have
been observed from neutron stars that take place in their
atmospheres, which is a very thin shell on their surfaces. In order to
predict observable quantities related to these phenomena we need to
parametrize only the values of the metric elements on their surfaces
and not throughout their exterior spacetimes. We can, therefore, use
observations to read directly different combinations of the metric
elements on the surfaces of the stars and compare them against general
relativistic predictions.

As an example, the atmospheric spectrum from a slowly spinning
bursting neutron star is determined by various metric elements that
depend on three parameters: the coordinate radius of the neutron-star
surface, the redshift from it, and the effective gravitational
acceleration at the same place. In general relativity all three
parameters depend only on the mass and radius of the neutron
star. Using, therefore, at least three atmospheric phenomena to
measure these three parameters independently can lead to a
quantitative test of this general relativistic prediction (Psaltis
2008).

\section*{Conclusions}

Testing general relativity in the strong-field regime may follow one
of two main approaches that are familiar from other areas of physics
or astrophysics. We can again draw a parallel between the directions
described in this article with current investigations of gravity at
cosmological scales. In one set of studies that aim to characterize
the accelerated expansion of the universe, the Friedman equation is
modified at a fundamental level by postulating, e.g., a scalar field
such as quintessence, and the parameters of the theory are constrained
against observations. In a different set of studies, the field that
provides the acceleration of the universe is described
phenomenologically by an equation of state, with parameters that can
be measured observationally. The results of these two sets of studies
offer a complimentary view of the observational data. At the same
time, the lessons learned and the interplay between the two approaches
allows for both to be further developed. The quality of high-energy
and of gravitational-wave data that are expected from future missions and
observatories require that both approaches be exploited as well in the 
case of strong-field tests of general relativity.

\section*{References}

\begin{thereferences}

\item{Arkani-Hamed N.\ Dimopoulos S.\ and Dvali G.\ 1998a {\em Phys.\ Lett.\ 
B} {\bf 429} 263}

\item{Arkani-Hamed N.\ Dimopoulos S.\ and Dvali G.\ 1998b {\em Phys.\ Rev.\ D.}
{\bf 59} 086044}

\item{Barausse E.\ and Sotiriou T.\ P.\ 2008 {\em Phys.\ Rev.\ Let.} {\bf 101}
9001}

\item{Burgess C.\,P.\ 2004 {\em Living Rev.\ Relativity} {\bf 7} lrr-2004-5}

\item{Chiba T.\ 2003 {\em Phys.\ Lett.\ B.} {\bf 575} 1}

\item{Collins N.\ A.\ and Hughes S.\ A.\ 2004 {\em Phys.\ Rev.\ D.} {\bf 69}
124022}

\item{Cooney A.\ DeDeo S.\ and Psaltis D.\ 2009 {\em Phys.\ Rev.\ D.} {\bf 79}
4033}

\item{Damour T.\ and Esposito-Farese G.\ 1993 {\em Phys.\ Rev.\ Let.}
{\bf 70} 2220}

\item{Dolgov A.\ D.\ and Kawasaki M.\ 2003 {\em Phys.\ Lett.\ B.} {\bf 573}
1}

\item{Eling C.\ and Jacobson T.\ 2006 {\em Class.\ Quantum Grav.} {\bf 23}
5642}

\item{Emparan R.\ Garcia-Bellido J.\ and Kaloper N.\ 2003 {\em High
Energy Phys.} {\bf 1} 79}

\item{Lanahan-Tremblay N.\ and Faraoni V.\ 2007 {\em Class.\ Quantum
Grav.} {\bf 24} 5667}

\item{Fitzpatrick A.\ L.\ Randall L.\ and Wiseman T.\ 2006 {\em JHEP}
{\bf 0611} 033}

\item{Flanagan E.\ E.\ and Hughes S.\ A.\ 2005 {\em New J.\ Phys.} {\bf 7} 
204}

\item{Glampedakis K.\ and Babak S.\ 2006 {\em Class.\ Quantum Grav.} 
{\bf 23} 4167}

\item{Hu W.\ and Sawicki I.\ 2007 {\em Phys.\ Rev.\ D.} {\bf 76} 4004}

\item{Hughes S.\ A.\ 2006 {\em AIP Conference Series} {\bf 873} 233}

\item{Johannsen T.\ Psaltis D.\ and McClintock J.\ 2009 {\em Astrophys.\ J.}
{\bf 691} 997}

\item{Kapner D.\ J.\ et al.\ 2007 {\em Phys.\ Rev.\ Let.} {\bf 98} 021101}

\item{Konno K.\ Matsuyama T.\ and Tanda S.\ 2009 {\em Preprint} 
arXiv:0902.4767}

\item{Kostelecky A.\ 2003 {\em Phys.\ Rev.\ D.} {\bf 69} 105009}

\item{Landau L.\ D.\ and Lifshitz E.\ M.\ 1980 {\em The Classical Theory
of Fields} (Butterworth-Heinemann; 5th edition)}

\item{Maartens R.\ 2005 {\em Living Rev.\ Relativity} {\bf 7}, lrr-2004-7}

\item{Magnano G.\ and Sokolowski 1994 {\em Phys.\ Rev.\ D} {\bf 50} 5039}

\item{Misner C.\ W. Thorne K.\ S. and Wheeler J.\ A.\ 1973 {\em Gravitation}
(W.H. Freeman: San Fransisco)}

\item{Nojiri S.\ and Odintsov S.\ D.\ 2003 {\em Phys.\ Lett.\ B} {\bf 576} 5}

\item{Psaltis D.\ 2009 {\em Living Rev.\ Relativity} {\bf 11}, lrr-2009-9}

\item{Psaltis D.\ Perrodin D.\ Dienes K.\ and Mocioiu I.\ 2008 {\em Phys.\
Rev.\ Let.} {\bf 100} 1101}

\item{Randall L.\ and Sundrum R. 1999 {\em Phys.\ Rev.\  Lett.} {\bf 83} 
4690}

\item{Ryan F.\ D.\ 1995 {\em Phys.\ Rev.\ D.} {\bf 52} 5707}

\item{Simon J.\ Z.\ 1990 {\em Phys.\ Rev.\ D.} {\bf 41} 3720}

\item{Sotiriou T.\ P.\ and Faraoni V.\ 2009 {\em Rev.\ Mod.\ Phys.}
{\em Preprint} arXiv:0805.1726}

\item{Stairs I.\ H.\ 2003 {\em Living Rev.\ Relativity} {\bf 6}, lrr-2003-5}

\item{Upadhye A.\ and Hu W.\ 2009 {\em Preprint} arXiv:0905.4055}

\item{Yunes N.\ and Pretorious F.\ 2009 {\em Phys.\ Rev.\ D.} {\em Preprint}
arXiv:0902.4669}

\item{Will C.\ M.\ 2006 {\em Living Rev.\ Relativity} {\bf 9}, lrr-2006-3}

\item{Woodard R.\ P.\ 2006 {\em Preprint} astro-ph/0601672}

\item{Zaglauer H.\ 1992 {\em Astrophys.\ J.} {\bf 393} 685}

\end{thereferences}

\end{document}